\documentclass{elsarticle}

\usepackage{hyperref}
\usepackage{latexsym}
\usepackage{theorem}
\usepackage{amsmath}
\usepackage{amssymb}
\usepackage{graphics}

\newcommand{\forsub}[1]{\ifthenelse{\equal{\version}{sub}}{#1}{}}
\newcommand{\forfull}[1]{\ifthenelse{\equal{\version}{full}}{#1}{}}
\newcommand{\forfinal}[1]{\ifthenelse{\equal{\version}{final}}{#1}{}}

\newcommand{\comment}[1]{}

\newcommand{\lex}{_{lex}}
\newcommand{\mul}{_{mul}}
\newcommand{\stat}[1][f]{_{stat_{#1}}}

\renewcommand{\a}{\rightarrow}
\newcommand{\A}{\Rightarrow}

\newcommand{\al}{\leftarrow}

\renewcommand{\to}{\mapsto}

\newcommand{\I}[1]{[\![#1]\!]}

\newcommand{\all}{\forall}
\newcommand{\ou}{\vee}
\newcommand{\et}{\wedge}

\newcommand{\sle}{\subseteq}

\newcommand{\tge}{\unrhd} 

\newcommand{\tgt}{\rhd}

\renewcommand{\b}{\beta}

\newcommand{\vep}{\varepsilon}
\renewcommand{\l}{\lambda}
\renewcommand{\L}{\Lambda}

\renewcommand{\S}{\Sigma}
\renewcommand{\t}{\theta}

\newcommand{\mi}{\mathit}
\newcommand{\mc}{\mathcal}
\newcommand{\mt}{\mathtt}
\newcommand{\mr}{\mathrm}

\newcommand{\cC}{\mc{C}}

\newcommand{\cF}{\mc{F}}

\newcommand{\cI}{\mc{I}}

\newcommand{\cP}{\mc{P}}

\newcommand{\cT}{\mc{T}}

\newcommand{\cX}{\mc{X}}

\newenvironment{rew}%
  {\begin{tabular}{r@{~~$\a$~~}l}}%
  {\end{tabular}}
\newenvironment{rewc}%
  {\begin{center}\begin{rew}}%
  {\end{rew}\end{center}}

\newenvironment{proof}%
  {{\em Proof. }}%
  {}

\newcounter{counter}

{\theorembodyfont{\rmfamily} 
  \newtheorem{defn}[counter]{Definition}
  \newtheorem{lem}[counter]{Lemma}
  \newtheorem{thm}[counter]{Theorem}

}

\begin{document}


\begin{frontmatter}

\title{Inductive Data Type Systems}

\comment{
\author{Fr\'ed\'eric Blanqui\and Jean-Pierre Jouannaud\\
%
LRI, B\^at. 490, Universit\'e Paris-Sud\\
91405 Orsay, FRANCE\\
Tel: (33) 1.69.15.69.05 \quad Fax: (33) 1.69.15.65.86\\
{\tt http://www.lri.fr/\~{}blanqui/}\\
%
\and Mitsuhiro Okada\\
%
Department of Philosophy, Keio University,\\
108 Minatoku, Tokyo, JAPAN}

\date{}
\maketitle
}

\author{Fr\'ed\'eric Blanqui, Jean-Pierre Jouannaud}

\address{LRI, B\^at. 490, Universit\'e Paris-Sud\\
91405 Orsay, FRANCE\\
Tel: (33) 1.69.15.69.05 \quad Fax: (33) 1.69.15.65.86\\
{\tt http://www.lri.fr/\~{}blanqui/}}

\author{Mitsuhiro Okada}

\address{Department of Philosophy, Keio University,\\
108 Minatoku, Tokyo, JAPAN}

\begin{abstract}
  In a previous work (``Abstract Data Type Systems'', TCS 173(2),
  1997), the last two authors presented a combined language made of a
  (strongly normalizing) algebraic rewrite system and a typed
  $\l$-calculus enriched by pattern-matching definitions following a
  certain format, called the ``General Schema'', which generalizes the
  usual recursor definitions for natural numbers and similar ``basic
  inductive types''. This combined language was shown to be strongly
  normalizing. The purpose of this paper is to reformulate and extend
  the General Schema in order to make it easily extensible, to capture
  a more general class of inductive types, called ``strictly
  positive'', and to ease the strong normalization proof of the
  resulting system. This result provides a computation model for the
  combination of an algebraic specification language based on abstract
  data types and of a strongly typed functional language with strictly
  positive inductive types.
\end{abstract}


\end{frontmatter}


\section{Introduction}

This work is one step in a long term program aiming at building formal
specification languages integrating computations and proofs within a
single framework. We focus here on incorporating an expressive notion
of equality within a typed $\l$-calculus.

In retrospect, the quest for an expressive language allowing to
specify and prove mathematical properties of software started with
system F on the one hand \cite{girard72thesis,girard88book} and the
Automath project on the other hand \cite{debruijn70}. Much later,
Coquand and Huet combined both calculi, resulting in the Calculus of
Constructions \cite{coquand85eurocal}. Making use of impredicativity,
data structures could be encoded in this calculus, but these encodings
were far too complex to be used by non-specialists. A different
approach was taken by Martin-L\"of \cite{martinlof73,martinlof84book},
whose theory was based on the notion of inductive definition,
originating in G\"odel's system T \cite{godel32}. Coquand and
Paulin-M\"ohring later incorporated a similar notion to the Calculus
of Constructions under the name of inductive type
\cite{coquand88colog}. But despite their legitimate success, inductive
types are not yet enough to make the Calculus of Inductive
Constructions an easy to use programming language for proofs. The main
remaining problem is that of equality.  In the current version of the
calculus, equality is given by $\b\eta$-reductions, the recursor rules
associated with the inductive types --corresponding to structural
induction in the Curry-Howard isomorphism--, and the definitional
rules for constants by primitive recursion of higher type. This notion
of equality has two main practical drawbacks: it makes the definition
of functions sometimes painful for the user, by forcing the user to
think operationally rather than axiomatically; it makes it necessary
to spell out many equational proofs that could be short-cutted if the
corresponding equality could be equationally specified in the
calculus.

It should be clear that this problem is not specific to the Calculus
of Inductive Constructions. It also shows up in other versions of type
theory where equality is not a first-class concept, for example, in
Martin Löf's theory of types. A solution was proposed by Coquand, for
a calculus with dependent types, in which functions can be defined by
pattern-matching, provided all right-hand side recursive calls are
``structurally smaller'' than the left-hand side call
\cite{coquand92types}. His notion is very abstract, though, and relies
on a well-foundedness assumption which is satisfied in practice.
Concurrently, following the pioneering works of Tannen
\cite{breazu88lics}, Tannen and Gallier
\cite{breazu89icalp,breazu91tcs} and Okada \cite{okada89issac}, the
last two authors of the present paper proposed another solution, for a
polymorphically typed $\l$-calculus, based on pattern-matching
functional definitions following the so-called ``General Schema''
\cite{jouannaud91lics,jouannaud97tcs}. This work was then generalized
so as to cover the full Calculus of Constructions
\cite{barbanera93tlca,barbanera93icalp,barbanera94lics}.  As in
Coquand \cite{coquand92types}, the idea of the General Schema is to
control the arguments of the right-hand side recursive calls of a
rule-based definition by checking that they are smaller than the
left-hand sides ones, this time in the strict subterm ordering
extended in a multiset or lexicographic manner. This schema was
general enough to subsume basic inductive types, such as $\mt{nat} =
0_\mt{nat} \uplus s_\mt{nat}(\mt{nat})$, in the sense that the
associated recursor rules are instances of the General Schema. In
contrast with Coquand's proposal, it does not subsume non-basic
inductive types, such as $\mt{ord} = 0_\mt{ord} \uplus
s_\mt{ord}(\mt{ord}) \uplus lim(\mt{nat} \a \mt{ord})$, whose
constructor $\lim$ takes an argument of the functional type $\mt{nat}
\a \mt{ord}$. On the other hand, the use of multiset and lexicographic
extensions allows to tailor the comparisons to the practical needs,
making it possible to have nested recursive calls, an important
facility that Coquand's ordering cannot provide with. Finally, it is
important to note that, in contrast with other work
\cite{mendler86tr,gimenez94types}, our definitions allow non-linear
and overlapping left-hand sides, to the price of checking confluence
via the computation of critical pairs.

The fact that the General Schema covers only a limited portion of the
possible inductive types of the Calculus of Inductive Constructions
shows a weakness, and indeed, functions defined by induction over such
inductive types cannot be defined within the schema. The purpose of
this paper is to revisit the General Schema so as to cover all
strictly positive inductive types. The solution is based on an
essential use of the positivity condition required for the inductive
types. We do so within the framework of Church's simple theory of
types, therefore avoiding the problem of having equalities at the type
level via the use of dependent types. Closing the gap between the
simple theory of types and the Calculus of Inductive Constructions
will require further generalizations of the General Schema allowing
for dependent and polymorphic inductive types.

The strong normalization proof of our new calculus is based on Tait's
computability predicates method \cite{tait67jsl,girard88book}. In
contrast with \cite{jouannaud97tcs}, the whole structure of the proof
is made quite modular thanks to a novel formulation of our new version
of the General Schema. Here, given a left-hand side $f(\vec{l})$, we
define the (infinite) set of possible right-hand sides $r$ such that
the rule $f(\vec{l}) \a r$ follows the schema. This set of right-hand
sides is generated inductively from $\vec{l}$ by computability
preserving operations. This new definition, as it can be easily seen,
is strictly stronger than the previous one, allows to reason by
induction on the construction of the set of possible right-hand sides,
and is easily extensible. This latter feature should prove very useful
when extending the present work to the Calculus of Inductive
Constructions.

We define our language in Section~\ref{sec-language}, ending with the
new definition of the General Schema in
Subsection~\ref{subsec-schema}. The normalization proof is given in
Section~\ref{sec-normalization}. In Section~\ref{sec-applications}, we
detail many examples and explain possible extension of the General
Schema in order to be able to prove some of them. We conclude in
Section~\ref{sec-conclusion} with two more, important open problems.


\section{Inductive Data Type Systems}
\label{sec-language}

Intuitively, an {\em Inductive Data Type System} (IDTS) is a
simply-typed $\l$-calculus in which each base type is equipped with a
set of constructors together with the associated structural induction
principle in the form of G\"odel's primitive recursive rules of higher
type and additional function symbols (completely) defined by
appropriate higher-order rewrite rules. The former kind of rules can
actually be seen as a particular case of the latter, resulting in a
uniform formalism with a strong rewriting flavor. In the sequel, we
assume the reader familiar with the notions of $\l$-calculus and term
rewriting, as presented in \cite{barendregt93book} for the
simply-typed $\l$-calculus, \cite{dershowitz90book} for term rewriting
and \cite{klop93tcs,nipkow91lics,oostrom94thesis} for the several
variants of higher-order rewriting existing in the literature.

We first introduce the term language before to move on with the
definition of higher-order rewrite rules and of the new formulation of
the General Schema.


\subsection{The language}

In this subsection, we introduce successively the signature (made of
inductive types, constructors and function symbols) and the set of
well-formed terms before to end up with the set of computational
rules.


\subsubsection{Signature}


\newcommand{\ts}{\mt{s}}

\begin{defn}[Types]
  Given a set $\cI$ whose elements are called {\em inductive types},
  the set $\cT$ of {\em types} is generated by the following grammar
  rule:

\begin{center}
$s = \ts ~|~ (s\a s)$
\end{center}

where $\ts$ ranges over $\cI$. Furthermore, we consider that $\a$
associates to the right, hence $s_1 \a (s_2 \a s_3)$ can be written
$s_1 \a s_2 \a s_3$.

The sets of positive and negative positions of a type $s$ are
inductively defined as follows:

\begin{center}
$Pos^+(s \in \cI) ~=~ \epsilon$\\
$Pos^-(s \in \cI) ~=~ \emptyset$\\
$Pos^+(s \a t) ~=~ 1 \!\cdot\! Pos^-(s) ~\cup~ 2 \!\cdot\! Pos^+(t)$\\
$Pos^-(s \a t) ~=~ 1 \!\cdot\! Pos^+(s) ~\cup~ 2 \!\cdot\! Pos^-(t)$
\end{center}

We say that an inductive type $\mt{t}$ occurs {\em positively} in a
type $s$ if $\mt{t}$ does occur in $s$ and every occurrence of
$\mt{t}$ in $s$ belongs to $Pos^+(s)$. $\mt{t}$ is said to occur {\em
  strictly positively} in $s_1 \a \ldots \a s_n \a \mt{t}$ if $\mt{t}$
occurs in no $s_i$.
\end{defn}

This notion of positivity/negativity associated to the type
constructor $\a$ is similar to the one used in logic with respect to
the implication operator $\A$ (as can be expected from the
Curry-Howard isomorphism). Note that if $\ts$ does not occur
positively in $t$ then, either $\ts$ does not occur in $t$ or else
$\ts$ occurs at a negative position in $t$. For example, $\mt{ord}$
occurs positively in $s = \mt{nat} \a \mt{ord}$ since it occurs in $s$
at the set of positive positions $\{1\} \sle Pos^+(s) = \{1\}$. In
fact, it does occur strictly positively since $\mt{ord}$ does not
occur in $\mt{nat}$. On the other hand, $\mt{ord}$ does not occur
positively in $t = \mt{ord} \a \mt{ord}$ since it occurs at the
negative position $1 \in Pos^-(t) = \{1\}$.


\begin{defn}[Constructors]
  We assume that each inductive type $\ts \in \cI$ comes along with an
  associated set $\cC(\ts)$ of {\em constructors}, each constructor $C
  \in \cC(\ts)$ being equipped with a type $\tau(C) = s_1 \a \ldots \a
  s_n \a \ts$. $n$ is called the {\em arity} of $C$ and we denote by
  $\cC^n$ the set of constructors of arity $n$. We assume that the
  sets $\cC(\ts)$ are pairwise disjoint.
  
  Constructor declarations define a quasi-ordering on $\cI$: an
  inductive type $\ts$ depends on an inductive type $\mt{t}$, written
  $\ts \ge_\cI \mt{t}$, if $\mt{t}$ occurs in the type of a
  constructor $C \in \cC(\ts)$. (In fact, we consider the reflexive
  and transitive closure of this relation.) We use $=_\cI$ and $>_\cI$
  for respectively the equivalence and the strict ordering associated
  to $\ge_\cI$ and say that $\ts$ is {\em $\cI$-equivalent} to
  $\mt{t}$ if $\ts =_\cI \mt{t}$.
\end{defn}

\begin{defn}[Strictly positive inductive types]
  An inductive type $\ts$ is said to be {\em strictly positive} if it
  does not occur or occurs strictly positively in the types of the
  arguments of its constructors, and no type $\cI$-equivalent to $\ts$
  occurs at a negative position in the types of the arguments of the
  constructors of $\ts$. A strictly positive type is {\em basic} if
  its constructors have no functional arguments.
\end{defn}

{\bf Assumption 1:} We assume that $>_\cI$ is well-founded and that
all inductive types are strictly positive.

To spell out the strict-positivity condition, assume that an inductive
type $\ts$ has $n$ constructors $C_1, \ldots, C_n$ with $\tau(C_i) =
s_{i,1} \a \ldots \a s_{i,n_i} \a \ts$ and $s_{i,j} = s_{i,j,1} \a
\ldots \a s_{i,j,n_{i,j}} \a \mt{t}_{i,j}$. Then, $\ts$ is strictly
positive if $\mt{t}_{i,j} \le_\cI \ts$, $\ts$ occurs in no
$s_{i,j,k}$ and no type $\cI$-equivalent to $\ts$ occurs at a negative
position in some $s_{i,j}$. It is basic if, moreover, $n_{i,j} = 0$
for all $i,j$.


Examples of type definitions used in the paper are $\mt{bool}$ for
booleans, $\mt{nat}$ for natural numbers, $\mt{list\_nat}$ for lists
of natural numbers (we do not consider polymorphic types here),
$\mt{tree}$ and $\mt{list\_tree}$ for the mutually inductive types of
trees and lists of trees, $\mt{proc}$ for process expressions
\cite{sellink93sosl} ($\delta$ denotes the deadlock, ``;'' the
sequencing, + the choice operator and $\S$ the dependent choice),
$\mt{ord}$ for well-founded trees, i.e. Brouwer's ordinals
\cite{stenlund72book}, $\mt{form}$ for formulas of the predicate
calculus and $\mt{R}$ for expressions built upon real numbers
\cite{prehofer95thesis}:

\begin{itemize}
\item $\mt{bool} = true:\mt{bool} ~|~ false:\mt{bool}$
\item $\mt{nat} = 0:\mt{nat} ~|~ s:\mt{nat} \a \mt{nat}$
\item $\mt{listnat} = nil:\mt{listnat} ~|~ cons:\mt{nat} \a
  \mt{listnat} \a \mt{listnat}$
\item $\mt{tree} = node:\mt{listtree} \a \mt{tree}$
\item $\mt{listtree} = nil:\mt{listtree} ~|~ cons:\mt{tree} \a
  \mt{listtree} \a \mt{listtree}$
\item $\mt{proc} = \delta:\mt{proc} ~|~ ;:\mt{proc} \a \mt{proc}
  \a \mt{proc} ~|~ +:\mt{proc} \a \mt{proc} \a \mt{proc} ~|~
  \S:(\mt{data} \a \mt{proc}) \a \mt{proc}$
\item $\mt{ord} = 0:\mt{ord} ~|~ s:\mt{ord} ~|~ lim:(\mt{nat} \a
  \mt{ord}) \a \mt{ord}$
\item $\mt{form} = \ou:\mt{form} \a \mt{form} \a \mt{form} ~|~
  \neg:\mt{form} \a \mt{form} ~|~ \all: (\mt{term} \a \mt{form})
  \a \mt{form} ~|~ \ldots$
\item $\mt{R} = 0:\mt{R} ~|~ 1:\mt{R} ~|~ +:\mt{R} \a \mt{R} \a
  \mt{R} ~|~ cos:\mt{R} \a \mt{R} ~|~ ln:\mt{R} \a \mt{R} ~|~
  \ldots$
\end{itemize}

All types above are basic, except $\mt{ord}$ and $\mt{form}$ which are
strictly positive. We have used the same name for constructors of
different types, but we should not if they have to live together.  For
the sake of simplicity, we will continue in practice to overload names
when there is no ambiguity, otherwise we will disambiguate names as in
$0_\mt{nat}$. Our inductive types above are inhabited by expressions
built up from their constructors, as for example $\all(\l x. (P ~x)
\et (Q ~x))$ which represents the logical formula $\all x\, P(x) \et
Q(x)$.


A more general class of inductive types is the one of {\em positive}
inductive types. An inductive type is said to be positive if it occurs
only at positive positions in the types of the arguments of its
constructors (the case of mutually inductive types is defined
similarly, by requiring that any type equivalent to it occurs only at
positive positions in the types of the arguments of its constructors).
The positivity condition ensures that we can define sets of objects by
induction on the structure of the elements of the inductive type: it
implies the monotonicity of the functional of which the set of objects
is the least fixpoint.  The class of positive inductive types is the
largest class that one can consider within the framework of the
simply-typed $\l$-calculus, since any non-positive type is inhabited
by non-terminating well-typed terms in this framework
\cite{mendler87thesis}. In this paper, we restrict ourselves to
strictly positive inductive types, as in the Calculus of Inductive
Constructions \cite{werner94thesis}, and prove the strong
normalization property of our calculus under this assumption. However,
we conjecture that strong normalization holds in the non-strictly
positive case too.


\begin{defn}[Function symbols]
  For each non empty sequence $s_1, \ldots, s_n,$ $s$ of types, we
  assume given a set $\cF_{s_1, \ldots, s_n, s}$ of {\em function
    symbols} containing the constructors of arity $n$ and type $s_1 \a
  \ldots \a s_n \a s$. Given a symbol $f \in \cF_{s_1, \ldots, s_n,
    s}$, $n$ is its {\em arity} and $\tau(f) = s_1 \a \ldots \a s_n \a
  s$ its {\em type}. We denote by $\cF^n$ the set of function symbols
  of arity $n$ and by $\cF$ the set of all function symbols.
  
  We also assume given a quasi-ordering $\ge_\cF$ on $\cF$, called
  {\em precedence}, whose associated strict ordering $>_\cF$ is
  well-founded.
\end{defn}

For example, we may have an injection function $i$ from $\mt{nat}$ to
$\mt{ord}$. Then, $lim(\l n. i(n))$ represents the first limit ordinal
$\omega$ as the limit of the infinite sequence of ordinals $0, s(0),
s(s(0)), \ldots$ We will later see how to define this injection
function in our calculus.


\subsubsection{Terms}

\begin{defn}[Terms]
\label{def-term}
Given a family $(\cX^s)_{s \in \cT}$ of disjoint infinite sets of {\em
  variables} with $\cX$ denoting their union, the set of {\em untyped
  terms} is defined by the grammar rule:

\begin{center}
  $u = x ~|~ \l x.u ~|~ (u ~u) ~|~ f(u_1, \ldots, u_n)$
\end{center}

\noindent
where $f$ ranges over $\cF^n$ and $x$ over $\cX$. $\l x. u$ denotes
the {\em abstraction} of $u$ w.r.t. $x$, i.e. the function of
parameter $x$ and body $u$, while $(u ~v)$ denotes the {\em
  application} of the function $u$ to the term $v$. A term of the form
$f(u_1, \ldots, u_n)$ is said to be {\em function-headed} and {\em
  constructor-headed} if $f \in \cC$.

The family of sets $(\L^s)_{s \in \cT}$ of {\em terms of type $s$} is
inductively defined on the structure of terms as follows:

\begin{itemize}
\item if $x\in \cX^s$ then $x\in \L^s$,
\item if $x\in \L^s$ and $u\in \L^t$ then $\l x.u \in \L^{s\a t}$,
\item if $u\in \L^{s\a t}$ and $v\in \L^s$ then $(u ~v) \in \L^t$,
\item if $f$ is a function symbol of arity $n$ and type $s_1 \a \ldots
  \a s_n \a s$ and $u_1 \in \L^{s_1}, \ldots, u_n \in \L^{s_n}$ then
  $f(u_1, \ldots, u_n) \in \L^s$.
\end{itemize}

\noindent
Finally, we denote by $\L = \bigcup_{s\in \cT} \L_s$ the set of {\em
  terms} of our calculus. The {\em type of a term $u$} is the (unique)
type $t \in \cT$ such that $u \in \L^t$. We may use the notation $u:t$
to indicate that $u$ is of type $t$.
\end{defn}


Note that we could have adopted a presentation based on type-checking
rules. The reader will easily extract such rules from the definition
of the sets $\L^s$.

As usual, we consider that the application associates to the left such
that $((u_1 ~u_2) ~u_3)$ can be written $(u_1 ~u_2 ~u_3)$. The
sequence of terms $u_1\ldots u_n$ is denoted by the vector $\vec{u}$
of length $|\vec{u}| = n$.  We consider that $(v ~\vec{u})$ and $\l
\vec{x}. v$ both denote the term $v$ if $\vec{u}$ or $\vec{x}$ is the
empty sequence, and the respective terms $(\ldots ((v ~u_1) ~u_2)
\ldots u_n)$ and $\l x_1 \ldots \l x_n. v$ otherwise.

After Dewey, the set $Pos(u)$ of {\em positions} in a term $u$ is a
language over the alphabet of strictly positive natural numbers. The
{\em subterm} of a term $u$ at position $p \in Pos(u)$ is denoted by
$u|_p$ and the term obtained by replacing $u|_p$ by a term $v$ is
written $u[v]_p$. We write $u\tge v$ if $v$ is a subterm of $u$.

We denote by $FV(u)$ the set of {\em free variables} occurring in a
term $u$. A term in which a variable $x$ occurs freely at most once is
said to be {\em linear w.r.t. $x$}, and a term is {\em linear} if all
its free variables are linear.

A {\em substitution} $\t$ is an application from $\cX$ to $\L$,
written in a postfix notation as in $x\t$. Its {\em domain} is the set
$dom(\t) = \{ x \in \cX ~|~ x\t \neq x \}$. A substitution is
naturally extended to an application from $\L$ to $\L$, by replacing
each free variable by its image and avoiding variable captures. This
can be carried out by renaming the bound variables if necessary, an
operation called {\em $\alpha$-conversion}.  As usual, we will always
work modulo $\alpha$-conversion, hence identifying the terms that only
differ from each other in their bound variables. Furthermore, we will
always assume that free and bound variables are distinct and that
bound variables are distinct from each other. Finally, we may use the
notation $\{ \vec{x} \to \vec{u} \}$ for denoting the substitution
which associates $u_i$ to $x_i$ for each $i$.


\subsubsection{Computational rules}

Our language is made of three ingredients: a typed $\l$-calculus, a
set of inductive types with their constructors and a set of function
symbols. As a consequence, there will be three kinds of rules in the
calculus: the two rules coming from the $\l$-calculus,

\begin{center}
\begin{tabular}{rcll}
$(\l x.u ~v)$ & $\a_\b$ & $u \{ x \to v \}$\\
$\l x.(u ~x)$ & $\a_\eta$ & $u$ & if $x \notin FV(u)$\\
\end{tabular}
\end{center}

the rules associated with the inductive types, for example:

\begin{rewc}
$natrec(X, Y, 0)$ & $X$\\
$natrec(X, Y, s(n))$ & $(Y ~n ~natrec(X, Y, n))$\\
\end{rewc}

for the inductive type $\mt{nat}$, and the rules used for defining the
function symbols, for example:

\begin{rewc}
$i(0_\mt{nat})$ & $0_\mt{ord}$\\
$i(s_\mt{nat}(x))$ & $s_\mt{ord}(i(x))$\\
\end{rewc}

for the injection function from $\mt{nat}$ to $\mt{ord}$. We can
immediately see that the recursor rules look very much like the rules
defining the injection. We will show in Section~\ref{sec-applications}
that the recursor rules for strictly positive inductive types follow
the General Schema defined in Subsection~\ref{subsec-schema} and,
therefore, the recursor rules need not be singled out in our technical
developments.


\subsection{Higher-order rewriting}
\label{subsec-hor}

Before to define the General Schema precisely, we need to introduce
the notion of higher-order rewriting that we use. Indeed, several
notions of higher-order rewriting exist in the literature. Ours is the
simplest possible: a term $u$ rewrites to a term $u'$ by using a rule
$l \a r$ if $u$ {\em matches} the left-hand side $l$ or, equivalently,
if $u$ is an {\em instance} of $l$ by some substitution $\t$. Matching
here is syntactic, that is, $u$ is $\alpha$-convertible to the
instance of $l$. In contrast, the more sophisticated notions of
higher-order rewriting defined by Klop (Combinatory Reduction Systems
\cite{klop80thesis,klop93tcs}), Nipkow (Higher-order Rewrite Systems
\cite{nipkow91lics,mayr98tcs}) and van Raamsdonk and van Oostrom
(Higher-Order Rewriting Systems
\cite{oostrom94thesis,raamsdonk96thesis}, generalizing both) are based
on higher-order pattern-matching, that is, $u$ must be
$\b\eta\alpha$-convertible to the instance of $l$.


\begin{defn}[Rewrite rules and rewriting]
\label{def-rew-rule}
A {\em rewrite rule} is a pair $l\a r$ of terms such that:

\begin{enumerate}
\item $l$ is headed by a function symbol,
\item $FV(r) \sle FV(l)$,
\item $l$ and $r$ have the same type.
\end{enumerate}

Given a set $R$ of rewrite rules, a term $u$ $R$-rewrites to a term
$u'$ at position $p\in Pos(u)$ with the rule $l\a r\in R$, written $u
\a^p_R u'$, if there exists a substitution $\t$ such that $u|_p = l\t$
and $u'= u[r\t]_p$.

The {\em defining rules} of a function symbol $f$ are the rules whose
left-hand side is headed by $f$.
\end{defn}

Condition (3) ensures that the reduction relation preserves types,
that is, $u$ and $u'$ have the same type if $u \a_R u'$, a property
called {\em subject reduction}.


We now give two more (classical) examples defining, for the first, the
(formal) addition on Brouwer's ordinals and, for the second, some
functions over lists. The first example is paradigmatic in its use of
strictly positive types which are not basic. The second example uses a
rule with an abstraction in the left-hand side. More complex examples
of the second kind will be given in Section~\ref{sec-applications}.

For the (formal) addition of Brouwer's ordinals,

\begin{rewc}
$x + 0$ & $x$\\
$x + s(y)$ & $s(x + y)$\\
$x + lim(F)$ & $lim(\l n. (x + (F ~n)))$\\
\end{rewc}

note that the first two rules are just a first-order ones, hence a
special case of higher-order rule. More important, note the need of an
abstraction in the right-hand side of the last rule to bind the
variable $n$ needed for using the higher-order variable $F$ taken from
the left-hand side. This makes the termination proof of this set of
rules a difficult task. In our case, the termination property will be
readily obtained by showing that these rules follow our (improved)
definition of the General Schema. The difficulty, of course, is simply
delegated to the strong normalization proof of the schema.

About Brouwer's ordinals \cite{stenlund72book}, note that only a
suitable choice of $F$'s provides a semantically correct ordinal
notation and that, for such a correct notation, the above formal
definition provides semantically correct ordinal addition.

For the functions overs lists,

\begin{rewc}
$append(nil,l)$ & $l$\\
$append(cons(x,l), l')$ & $cons(x, append(l, l'))$\\
$append(append(l, l'), l'')$ & $append(l, append(l', l''))$\\
\end{rewc}

\begin{rewc}
$map(F, nil)$ & $nil$\\
$map(F, cons(x,l))$ & $cons((F ~x), map(F, l))$\\
$map(F, append(l, l'))$ & $append(map(F, l), map(F, l'))$\\
$map(\l x. x, l)$ & $l$\\
\end{rewc}

note that the three first rules, which define the concatenation
$append$ of two lists, are again usual first-order rules. The four
next rules define the function $map$ which successively applies the
function $F$ to the elements of some list. Note that the third and
sixth rule use a matching over a function symbol, namely $append$.


\subsection{The General Schema}
\label{subsec-schema}

We now proceed to describe the schema that the user-defined
higher-order rules should follow. In particular, all examples of
higher-order rules given so far satisfy this schema. It is inspired
from the last two authors former General Schema
\cite{jouannaud91lics,jouannaud97tcs} although the formulation is
quite different. The new schema is more powerful and answers a problem
left open with the former one, that is, the ability of capturing
definitions like the one previously given for the addition on
ordinals. The main property of the schema is that it ensures the
termination property of the relation $\a_R \cup \a_{\b\eta}$, for any
set $R$ of rules following the General Schema. This will be the
subject of Section~\ref{sec-normalization}.

In a function definition, in the case of a recursive call, we need a
way to compare the arguments of the recursive calls in the right-hand
side with the arguments of the left-hand side, and prove that they
strictly decrease to ensure termination. What we expect to use as the
comparison ordering is the subterm ordering or some extension of it.
The one we are going to introduce is similar to Coquand's notion of
``structurally smaller'' \cite{coquand92types} and will allow us to
deal with definitions like the addition on ordinals. The comparison
between the recursive call arguments and the left-hand side arguments
will then be done in a lexicographic or multiset manner, or a
combination thereof, according to a {\em status} of the function
symbol being defined. This status can be given by the user, or
computed in non-deterministic linear time.


In the following, we assume given a family $\{ x_i \}_{i \ge 1}$ of
variables.

\begin{defn}[Status ordering]
\label{def-status}
A {\em status} is a linear term $stat = lex(u_1$, \ldots, $u_p)$ $(p
\ge 1)$ where each $u_i$ is of the form $mul(x_{k_1}, \ldots,
x_{k_q})$ $(q \ge 1)$ with $x_{k_1}$, \ldots, $x_{k_q}$ of the same
type. The {\em arity} of $stat$ is the greatest indice $i$ such that
$x_i$ occurs in $stat$. The set $Lex(stat)$ of {\em lexicographic
  positions} in $stat$ is the set of indices $i$ such that there
exists $j \in \{ 1, \ldots, p \}$ for which $u_j = mul(x_i)$, that is,
$q = 1$.

Given a status $stat$ of arity $n$, a strict ordering $>$ on a set $E$
can be extended to an ordering $>\stat[]$ on sequences of elements of
$E$ of length greater or equal to $n$ as follows:

\begin{itemize}
\item $\vec{u} >\stat[] \vec{v}$ ~iff~ $stat \{ \vec{x} \to \vec{u} \}
  >\stat[]^{lex} stat \{ \vec{x} \to \vec{v} \}$
\item $lex(\vec{u}) >\stat[]^{lex} lex(\vec{v})$ ~iff~ $\vec{u}
  (>\stat[]^{mul})\lex \vec{v}$
\item $mul(\vec{u}) >\stat[]^{mul} mul(\vec{v})$ ~iff~ $\{ \vec{u} \}
  >\mul \{ \vec{v} \}$
\end{itemize}

where $>\lex$ and $>\mul$ denote the lexicographic and multiset
extension of $>$ respectively.
\end{defn}

For example, with $stat = lex(x_3, mul(x_2, x_4))$, $\vec{u}
\tgt\stat[] \vec{v}$ iff $u_3 \tgt v_3$ or else $u_3 = v_3$ and
$\{u_2, u_4\} \tgt\mul \{v_2, v_4\}$. Note that a status ordering
$stat$ boils down to the usual lexicographic ordering if $stat =
lex((mul(x_1), \ldots, mul(x_n))$ or to the multiset ordering if $stat
= lex(mul(x_1, \ldots, x_n))$. An important property of status
orderings is that $>\stat[]$ is well-founded if $>$ is well-founded.


The notion of status will allow us to accept definitions like the ones
below. For the Ackermann function $Ack$, we need to take the
lexicographic status $stat_{Ack} = lex(mul(x_1), mul(x_2))$ and, for
the binomial function $Bin(n, m) = C^n_{m+n}$, we need to take the
multiset status $stat_{Bin} = lex(mul(x_1, x_2))$.

\begin{rewc}
$Ack(0,y)$ & $s(y)$\\
$Ack(s(x), 0)$ & $Ack(x, s(0))$\\
$Ack(s(x), s(y))$ & $Ack(x, Ack(s(x), y))$\\
\end{rewc}

\begin{rewc}
$Bin(0, m)$ & $s(0)$\\
$Bin(s(n), 0)$ & $s(0)$\\
$Bin(s(n), s(m))$ & $Bin(n, s(m)) + Bin(s(n), m)$\\
\end{rewc}

Apart from the notion of status, the other ingredients of our schema
are new. We introduce them in turn.


\begin{defn}[Symbol definitions]
  We assume that each function symbol $f$ of arity $n \ge 1$ comes
  along with a status $stat_f$ of arity $p$ such that $1 \le p \le n$
  and a set $R_f$ of rewrite rules defining $f$. We denote by $R$ the
  set of all rewrite rules and by $\a \,=\, \a_R \cup \a_{\b\eta}$ the
  rewrite relation of the calculus.
\end{defn}

{\bf Assumption 2:} We assume that the precedence $>_\cF$ is
well-founded and that $stat_f = stat_g$ whenever $f =_\cF g$.


The main new idea in the definition of the General Schema is to
construct a set of admissible right-hand sides, once a left-hand side
is given. This set will be generated inductively from a starting set
of terms extracted from the left-hand side, called the set of {\em
  accessible} subterms, by the use of {\em computability} preserving
operations. Here, computability refers to Tait's computability
predicate method for proving the termination of the simply-typed
$\l$-calculus \cite{tait67jsl}, which was later extended by Girard to
the polymorphic $\l$-calculus \cite{girard71,girard88book}.

To explain our construction, we need to recall the basics of Tait's
method. The starting observation is that it is not possible to prove
the termination of $\b$-reduction directly by induction on the
structure of terms because of the application case: in the untyped
$\l$-calculus, the term $(\l x. xx ~\l x. xx)$ rewrites to itself
although $\l x. xx$ is in normal form.  Tait's idea was to strengthen
the induction hypothesis by using instead a property, the {\em
  computability}, implying termination. The computability predicate
can be defined by induction on the type of terms as follows: for an
inductive type $\ts$, take $\I{\ts} = SN^\ts$, the set of strongly
normalizable terms of type $\ts$ (terms having no infinite sequence of
rewrites issued from them). For a functional type $s\a t$, take $\I{s
  \a t} = \{ u \in \L^{s\a t} ~|~ \all v \in \I{s}, ~(u ~v) \in \I{t}
\}$. From this definition, it is easy to prove that every computable
term is strongly normalizable ($\I{s} \sle SN^s$) and that every term
is computable ($\L^s \sle \I{s}$). Therefore, every term is strongly
normalizable. The role of the General Schema when rewrite rules are
added is to ensure that computability is preserved along the added
rewritings.  This is why we require that a right-hand side of rule is
built up from subterms of the left-hand side, the {\em accessible}
ones, by computability preserving operations: a set called the {\em
  computable closure} of the left-hand side.


\begin{defn}[Accessible subterms]
Given a term $v$, the set $Acc(v)$ of {\em accessible} subterms of $v$
is inductively defined as follows:

\begin{enumerate}
\item $v \in Acc(v)$,
\item if $\l x.u \in Acc(v)$ then $u \in Acc(v)$,
\item if $C(\vec{u}) \in Acc(v)$ then each $u_i \in Acc(v)$,
\item if $(u ~x) \in Acc(v)$ and $x \notin FV(u) \cup FV(v)$ then $u
  \in Acc(v)$,
\item if $u$ is a subterm of $v$ of basic type such that $Fv(u) \sle
  FV(v)$ then $u\in Acc(v)$.
\end{enumerate}
\end{defn}

To see how this works, let us consider the examples of $append$ and
$map$ given in Subsection~\ref{subsec-hor}. For the rule $append(nil,
l) \a l$, $l$ is accessible in the arguments of $append$ by (1). For
the rule $append(cons(x, l), l') \a cons(x, append(l, l'))$, $l$ is
accessible in $cons(x, l)$ by (3) and (1). The other rules are dealt
with in the same way. Another example is given by the associativity
rule of the addition on natural numbers: in the rule $(x+y)+z \a
x+(y+z)$, the variables $x$ and $y$ are accessible by (5). This does
not work for the addition on Brouwer's ordinals since $\mt{ord}$ is
not a basic inductive type. The cases (2) and (4) will be useful in
the more complex examples of Section~\ref{sec-applications}.


We have already seen how to extract subterms from a left-hand side of
rule. We are left with the construction of the computable closure from
these subterms. Among the operations used for the computable closure,
one constructs recursive calls with ``smaller'' arguments. We
therefore need to define the intended ordering, which has to be richer
than the usual subterm ordering as examplified by the last rule of the
definition of the addition on Brouwer's ordinals:

\begin{rewc}
$x + lim(F)$ & $lim(\l n. (x + (F ~n)))$\\
\end{rewc}

We see that $(F ~n)$, the second argument of the recursive call, is
not a strict subterm of $lim(F)$. Extending the General Schema so as
to capture such definitions was among the open problems mentioned in
\cite{jouannaud97tcs}. On the other hand, in a set-theoretic
interpretation of functions as input-output pairs, the pair $(n, (F
~n))$ would belong to $F$, and therefore, $(F ~n)$ would in this sense
be smaller than $F$. This is what is done by Coquand with his notion
of ``structurally smaller'' \cite{coquand92types} which he assumes to
be well-founded without a proof. Here, we make the same idea more
concrete by relating it to the strict positivity condition of
inductive types.


\begin{defn}[Ordering on arguments]
  Let $s$ be a type and $u$ and $v$ be two terms of type $s$.
\begin{itemize}
\item If $s$ is a strictly positive inductive type then $u$ is {\em
    greater than\,} $v$, $u > v$, if there is $p\in Pos(u)$ such that
  $p\neq\vep$, $v=(u|_p ~\vec{v})$ and, for all $q< p$, $u|_q$ is
  constructor-headed.
\item Otherwise, $u > v$ if $v$ is a strict subterm of $u$ such that
  $FV(v) \sle FV(u)$.
\end{itemize}
\end{defn}


We are now ready to define the {\em computable closure} of a left-hand
side.

\newcommand{\C}{\cC\cC}

\begin{defn}[Computable closure]
\label{ccd}
Given a symbol $f \in \cF_{s_1, \ldots, s_n, s}$, the {\em computable
  closure} $\C_f(\vec{l})$ of some term $f(\vec{l})$ is inductively
defined as the least set $\C$ such that:

\begin{enumerate}
\item if $x$ is a variable then $x \in \C$,
\item if $u \in Acc(\vec{l})$ then $u \in \C$,
\item if $u$ and $v$ are two terms in $\C$ of respective types $t_1\a
  t_2$ and $t_1$ then $(u ~v) \in \C$,
\item if $u\in \C$ then $\l x. u\in \C$,
\item if $g \in \cF_{t_1, \ldots, t_p, t}$, $g <_\cF f$ and $u_1,
  \ldots, u_p$ are $p$ terms in $\C$ of respective types $t_1, \ldots,
  t_p$ then $g(\vec{u}) \in \C$,
\item if $g \in \cF_{t_1, \ldots, t_p, t}$, $g =_\cF f$ and $u_1,
  \ldots, u_p$ are $p$ terms in $\C$ of respective types $t_1, \ldots,
  t_p$ then $g(\vec{u}) \in \C$ whenever:
\begin{itemize}
\item $\vec{l} >\stat \vec{u}$,
\item if $l_i > (l_i|_p ~\vec{v})$ then each $v_i$ belongs to $\C$.
\end{itemize}
\end{enumerate}
\end{defn}


\begin{defn}[General Schema]
  A rewrite rule $f(\vec{l}) \a r$ follows the {\em General Schema
    (GS)} if $r \in \C_f(\vec{l})$ and, for every $x\in FV(r)$, $x\in
  Acc(\vec{l})$.
\end{defn}

As an example, let us prove that the definitions of $append$ and $map$
given in Subsection~\ref{subsec-hor} indeed follow the General Schema.
We already saw that the free variables occuring in the left-hand sides
were all accessible hence, by (2), they belong to the computable
closure (CC) of their respective left-hand side. For the rule
$append(cons(x, l), l') \a cons(x, append(l, l'))$, $append(l, l')$
belongs to (CC) by (6) since $l$ is a strict subterm of $cons(x, l)$.
For the rule $map(F, cons(x, l)) \a cons((F ~x), map(F, l))$, $(F ~x)$
belongs to (CC) by (3), $map(F, l)$ by (6) and the whole right-hand
side by (5). The other rules are dealt similarly.

In our previous definition of the General Schema, the computable
closure was kind of implicit with, in particular, a poor accessibility
relation and a case $(7)$ in which the ordering used was always the
strict subterm ordering.


The main differences with Coquand's notion of ``structurally smaller''
\cite{coquand92types} or its extension by Gim\'enez
\cite{gimenez94types} are that:

\begin{enumerate}
\item we use statuses for comparing the arguments of the recursive
  calls with the left-hand side arguments (which include lexicographic
  comparisons),
\item we may compare a function-headed term or a $\l$-headed term with
  one of its subterm while, in Coquand's definition, comparisons are
  restricted to constructor-headed terms.
\end{enumerate}

A main advantage of both notions of accessibility and computable
closure is their formulation: it is immediate to add new cases in
these definitions. This flexibility should of course be essential when
extending the schema to richer calculi.

Given a user's specification following the General Schema, the
question arises whether the following properties are satisfied:
subject reduction, confluence, completeness of definitions and strong
normalization. Subject reduction follows easily. Confluence reduces to
local confluence once strong normalization is satisfied and can
therefore be checked on the critical pairs. Completeness of
definitions is necessary for the recursor definitions to make sense in
our Curry-Howard interpretation of types. Checking it can be done by
solving (higher-order) disequations. As recalled in
\cite{jouannaud97tcs}, this can be done automatically for a reasonable
fragment of the set of second order terms. In the next section, we
address the remaining problem, strong normalization.



\section{Strong normalization}
\label{sec-normalization}

In this section, we prove that the rewrite relation $\a \,=\, \a_R
\cup \a_{\b\eta}$ is terminating, i.e. there is no infinite sequence
of rewrites, whenever all rules of $R$ satisfy the General Schema.
Due to the formulation of the schema, our proof here is much simpler
than the one in \cite{jouannaud97tcs}, although the schema is more
general. It is again based on Tait's computability predicate method.
See \cite{gallier90book} for a comprehensive survey of the method.

We first define the interpretation of types and prove important
properties about it. In a second part, we prove a computability
property for the function symbols: assuming that the rules satisfy the
General Schema, a term headed by a function symbol is computable
whenever its arguments are computable. Strong normalization follows
then easily.

 
\subsection{Interpretation of types}

\begin{defn}[Interpretation of types]
  The interpretation $\I{s}$ of a type $s \in \cT$ is inductively
  defined as follows:

\begin{itemize}
\item $\I{\ts \in \cI}$ is the set of terms $u \in SN^\ts$ such that,
  for all term $C(\vec{u})$ such that $u \a^* C(\vec{u})$, each $u_i
  \in \I{s_i}$,
\item $\I{s \a t} = \{ u \in \L^{s\a t} ~|~ \all v \in \I{s}, ~(u ~v)
  \in \I{t} \}$.
\end{itemize}

In the following, we will say that a term of type $s$ is {\em
  computable} if it belongs to $\I{s}$ and that a substitution $\t$ is
{\em computable} if, for every variable $x \in dom(\t) \cap \cX^s$,
$x\t \in \I{s}$.
\end{defn}

The reason why we need such a complex interpretation is because we
need the property that the arguments of a computable
constructor-headed term are computable. Meanwhile, we will see in
Lemma~\ref{lem-comp-prop}.7 just below that, in case of a basic
inductive type $\ts$, the interpretation is merely $SN^\ts$.


But, first, we show that our definition makes sense.

\begin{lem}
For every type $s\in\cT$, $\I{s}$ is uniquely defined.
\end{lem}

\begin{proof}
  It suffices to prove that it holds for every inductive type $\ts \in
  \cI$. For the sake of simplicity, we assume that $=_\cI$ is the
  identity, that is, there is no mutually inductive types. At the end,
  we tell how to treat the general case which, apart from the
  notations, is no more difficult. Let $\cP(SN^\ts)$ be the set of
  subsets of $SN^\ts$. $\cP(SN^\ts)$ is a complete lattice with
  respect to set inclusion $\sle$. We show that $\I{\ts}$ is uniquely
  defined as the least fixpoint of a monotone functional over this
  lattice. The proof is by induction on $>_\cI$ which is assumed to be
  well-founded.
  
  We define the following family of functions $F_\ts: \cP(SN^\ts) \a
  \cP(SN^\ts)$ indexed by inductive types:

$F_\ts(X) = X ~\cup ~\left\{
             u\in SN^\ts \left|
\left.\begin{array}{l}
     \text{if} ~u \a^* C(\vec{u}) ~\text{then each} ~u_i \in R_{s_i}(X)\\
      \end{array}
\right.
                         \right.\right\}$,\\
where
$R_t(X) = 
  \left\{
    \begin{array}{l@{~~\mr{if}~~}l}
      \I{\mt{t}} & t = \mt{t} \in \cI ~\mr{and}~ \ts >_\cI \mt{t}\\
      X & t=\ts\\
      R_{t_1}(X) \a R_{t_2}(X) & t = t_1 \a t_2\\
    \end{array}
   \right.$

Since inductive types are assumed to be (strictly) positive, $F_\ts$
is monotone. Hence, from Tarski's theorem, it has a least fixed point,
$\I{\ts}$.

In case of mutually inductive types, the function $F_\ts$ operates on
a product of subsets of $SN^{\ts_1} \times \ldots \times SN^{\ts_n}$
if $\ts_1, \ldots, \ts_n$ are all the inductive types equivalent to
$\ts$, which is again a lattice. Apart from the notations, the
argument is therefore the same.
\end{proof}


We showed that each $\I{\ts}$ is the least fixpoint of the monotone
functional $F_\ts$. This least fixpoint can be reached by transfinite
iteration. Let $F_\ts^\frak{a}$ be the $\frak{a}$-th iterate of
$F_\ts$ over the empty set. Note that we need to go further than
$\omega$ as it is shown by the following example. Consider the
function $f:\mt{nat}\a\mt{ord}$ defined by the following rules:

\begin{rewc}
$f(0_\mt{nat})$ & $0_\mt{ord}$\\
$f(s_\mt{nat}(n))$ & $lim(\l x. f(n))$\\
\end{rewc}

For all $n$, $f(n) \in F_\mt{ord}^{n+1} \setminus F_\mt{ord}^n$. Thus,
$lim(\l x.f(x)) \in F_\mt{ord}^{\omega+1} \setminus
F_\mt{ord}^\omega$.

This provides us a well-founded ordering on the computable terms of
type $\ts$:

\begin{defn}[Ordering on the arguments of a function symbol]
  The {\em order\,} of a term $t \in \I{\ts}$ is the smallest ordinal
  $\frak{a}$ such that $t \in F_\ts^\frak{a}$. We say that $t \in
  \I{s}$ is {\em greater than\,} $u \in \I{s}$, $t \succ u$, if:

\begin{itemize}
\item $s\in \cI$ and the order of $t$ is greater than the order of
  $u$,
\item $s=s_1\a s_2$ and $t \a\!\cup\!\tgt u$.
\end{itemize}
\end{defn}

This is this ordering which allows us to treat the definitions on
strictly positive types. This idea is already used by Mendler for
proving the strong normalization of System F with recursors for
positive inductive types \cite{mendler87thesis} and by Werner for
proving the strong normalization of the Calculus of Inductive
Constructions with recursors for strictly positive types
\cite{werner94thesis}. We apply this technique to a larger class of
higher-order rewrite rules.

Let us see the example of the addition on Brouwer's ordinals. If
$lim(f)$ is computable then, by definition of $\I{\mt{ord}}$, $f$ is
computable. This means that, for any $n\in \I{\mt{nat}}$, $(f~n)$ is
computable. Therefore, $lim(f) \succ (f~n)$.


\begin{lem}[Computability properties]
\label{lem-comp-prop}
A term is {\em neutral} if it is not an abstraction nor
constructor-headed.

\begin{enumerate}
\item Every computable term is strongly normalizable.
\item Every strongly normalizable term of the form $(x ~\vec{u})$ is
  computable.
\item A neutral term is computable if all its immediate reducts are
  computable.
\item $(\l x. u ~v)$ is computable if $v$ is strongly normalizable and
  $u \{ x \to v \}$ is computable.
\item A constructor-headed term $C(\vec{u})$ is computable if the
  terms in $\vec{u}$ and all the immediate reducts of $C(\vec{u})$ are
  computable.
\item Computability is preserved by reduction.
\item If $\ts \in \cI$ is a basic inductive type then $\I{\ts} =
  SN^\ts$.
\end{enumerate}
\end{lem}

\begin{proof}
\begin{description}

\item [(1) and (2)] are proved together by induction on the type $s$ of
  the term.

\begin{description}

\item [$s = \ts \in \cI$:]

\hfill
\begin{enumerate}
\item $\I{\ts} \sle SN^\ts$ by definition.
\item Every strongly normalizable term $(x ~\vec{u})$ of type $\ts$ is
  computable since it cannot reduce to a constructor-headed term.
\end{enumerate}

\item [$s = s_1 \a s_2$:]

\hfill
\begin{enumerate}
\item Let $u$ be a computable term of type $s$ and $x$ be a variable
  of type $s_1$. By induction hypothesis, $x \in \I{s_1}$ hence, by
  definition of the interpretation for $s$, $(u ~x) \in \I{s_2}$. By
  induction hypothesis again, $(u ~x) \in SN^{s_2}$. Therefore, $u \in
  SN^{s_1 \a s_2}$.
\item Let $(x ~\vec{u})$ be a strongly normalizable term of type $s$
  and let $v \in \I{s_1}$. By induction hypothesis, $v \in SN^{s_1}$
  and $(x ~\vec{u} ~v) \in \I{s_2}$. Therefore, $(x ~\vec{u}) \in
  \I{s_1}$.
\end{enumerate}

\end{description}

\item [(3)] is proved again by induction on the type $s$ of the term.
  
\begin{description}
  
\item [$s = \ts \in \cI$:]

\hfill\\
  Let $u$ be a neutral term of type $\ts$ whose immediate reducts
  belong to $\I{\ts}$. By (1), its immediate reducts are strongly
  normalizable, hence $u \in SN^\ts$. Suppose now that $u$ reduces to
  a constructor-headed term $C(\vec{v})$. Since $u$ is neutral, it
  cannot be itself constructor-headed. Hence, $C(\vec{v})$ is a reduct
  of some immediate reducts $u'$ of $u$. By definition of $\ts$ and
  since $u' \in \I{\ts}$ by assumption, the terms in $\vec{v}$ are
  computable. Therefore $u \in \I{\ts}$.
  
\item [$s = s_1 \a s_2$:]

\hfill\\  
  Let $u$ be a neutral term of type $s$ whose immediate reducts are
  computable and let $v \in \I{s_1}$. By (1), $v \in SN^{s_1}$.
  Therefore, $\a$ is well-founded on the set of reducts of $v$.
    
  Then, we prove that the immediate reducts of $(u ~v)$ belong to
  $\I{s_2}$, by induction on $v$ w.r.t. $\a$. As $u$ is neutral, an
  immediate reduct of $(u ~v)$ is either of the form $(u' ~v)$ where
  $u'$ is a reduct of $u$, or else of the form $(u ~v')$ where $v'$ is
  a reduct of $v$. In the first case, since $u'$ is computable by
  assumption, $(u' ~v) \in \I{s_2}$. In the second case, we conclude
  by induction hypothesis on $v'$.
    
  As a consequence, since $(u ~v)$ is neutral, by induction
  hypothesis, $(u ~v) \in \I{s_2}$. Therefore, $u$ is computable.

\end{description}

\item [(4)] Since $(\l x. u ~v)$ is neutral, by (3), it suffices to
  prove that each one of its reducts is computable. The reduct $u \{ x
  \to v \}$ is computable by assumption. Otherwise, we reason by
  induction on the set of the reducts of $u$ and $v$ (which are both
  strongly normalizable) with $\a$ as well-founded ordering.
  
\item [(5)] Let $C(\vec{u})$ be a constructor-headed term such that
  the terms in $\vec{u}$ and all its immediate reducts are computable.
  Then, it is strongly normalizable since, by (1), all its immediate
  reducts are strongly normalizable. Now, let $D(\vec{v})$ be a
  constructor-headed term such that $C(\vec{u}) \a^* D(\vec{v})$. If
  $D(\vec{v}) = C(\vec{u})$ then the terms in $\vec{v} = \vec{u}$ are
  computable by assumption. Otherwise, there is an immediate reduct
  $v$ of $C(\vec{u})$ such that $v \a^* D(\vec{v})$. Since, by
  assumption, $v$ is computable, the terms in $\vec{v}$ are
  computable. Hence, $C(\vec{u})$ is computable.

\item [(6)] is proved again by induction on the type $s$ of the term.
  
\begin{description}
  
\item [$s = \ts \in \cI$:] 

\hfill\\  
  Let $u \in \I{\ts}$ and $u'$ be a reduct of $u$. By (1), $u \in
  SN^\ts$, hence $u' \in SN^\ts$. Besides, if $u'$ reduces to a
  constructor-headed term $C(\vec{v})$ then $u$ reduces to
  $C(\vec{v})$ as well. Therefore, by definition of $\I{\ts}$, the
  terms in $\vec{v}$ are computable and $u' \in \I{\ts}$.
  
\item [$s = s_1 \a s_2$:]

\hfill\\  
  Let $u$ be a computable term of type $s$, $u'$ be a reduct of $u$
  and $v \in \I{s_1}$. $(u' ~v)$ is a reduct of $(u ~v)$ which, by
  definition of $\I{s}$, belongs to $\I{s_2}$.  Hence, by induction
  hypothesis, $(u' ~v) \in \I{s_2}$ and $u'$ is computable.

\end{description}

\item [(7)] By (1), $\I{\ts} \sle SN^\ts$. We prove that $SN^\ts \sle
  \I{\ts}$, by induction on $SN^\ts$ with $\a \cup ~\tgt$ as
  well-founded ordering. Let $u \in SN^\ts$ and suppose that $u \a^*
  C(\vec{v})$ where $C \in \cC(\ts)$. Since $\ts$ is basic, $\tau(C) =
  \ts_1 \a \ldots \a \ts_n \a \ts$ where each $\ts_i$ is also a basic
  inductive type. Each $v_i$ is strongly normalizable hence, by
  induction hypothesis, each $v_i$ is computable. Therefore, $u$ is
  computable.

\end{description}
\end{proof}


\subsection{Computability of function symbols}

We start this paragraph by proving that accessibility is compatible
with computability, that is, any term accessible in a computable term
is computable. Then, we prove the same property for the computable
closure.


\begin{lem}[Compatibility of accessibility with computability]
\label{lem-comp-acc-red}
Let $v$ be a term and $\t$ a computable substitution such that
$dom(\t) \sle FV(v)$ and $v\t$ is computable. If $u$ is accessible in
$v$ and $\t'$ is a computable substitution such that $dom(\t') \cap
FV(v) = \emptyset$ then $u\t\t'$ is computable.
\end{lem}

\begin{proof}
 By induction on $u \in Acc(v)$.

\begin{enumerate}
\item The case $u = v$ is immediate since $u\t\t' = v\t\t' = v\t$.
  
\item $\l x. u \in Acc(v)$. $\t' = \t'' \uplus \{ x \to x\t' \}$ with
  $x \notin dom(\t'')$. $u\t\t' = u\t\t'' \{ x \to x\t' \}$ is a
  reduct of $(\l x. u\t\t'' ~x\t')$. $dom(\t'') \cap FV(v) =
  \emptyset$ hence, by induction hypothesis, $\l x.  u\t\t''$ is
  computable. Therefore, $u\t\t'$ is computable since, by assumption
  on $\t'$, $x\t'$ is computable.
  
\item $u = u_i$ and $C(\vec{u}) \in Acc(v)$. By induction hypothesis,
  $C(\vec{u}\t\t')$ is computable. Therefore, by definition of
  computability for inductive types, $u\t\t'$ is computable.
  
\item $(u ~x) \in Acc(v)$ and $x \notin FV(u) \cup FV(v)$. $u$ must be
  of type $s \a t$ and $x \notin dom(\t')$. Then, let $w$ be a
  computable term of type $s$ and $\t'' = \t' \uplus \{ x \to w \}$.
  $dom(\t'') \cap FV(v) = \emptyset$ hence, by induction hypothesis,
  $(u ~x)\t\t'' = (u\t\t' ~w)$ is computable. Therefore $u\t\t'$ is
  computable.
  
\item $u$ is a subterm of $v$ of basic type such that $FV(u) \sle
  FV(v)$. Since $FV(u) \sle FV(v)$, $u\t\t'=u\t$ is a subterm of
  $v\t$. Since $v\t$ is computable, hence strongly normalizable, its
  subterm $u\t$ is also strongly normalizable, hence computable, since
  it is of basic type.
\end{enumerate}
\end{proof}


\begin{lem}[Computability of function symbols]
\label{lem-comp-hofs}
Assume that the rules of $R$ satisfy the General Schema. For every
function symbol $f$, if $f(\vec{u})$ is a term whose arguments are
computable, then $f(\vec{u})$ is computable.
\end{lem}

\begin{proof}
  The proof uses three levels of induction: on the function symbols
  ordered by $>_\cF$ (H1), on the arguments of $f$ (H2) and on the
  right-hand side structure of the rules defining $f$ (H3).
  
  After Lemma~\ref{lem-comp-prop}.3 and \ref{lem-comp-prop}.5 (the
  terms in $\vec{u}$ are computable by assumption), $f(\vec{u})$ is
  computable if all its immediate reducts $w$ are computable. We prove
  that by induction on $(\vec{u}, \vec{u})$ with $(\succ\stat,
  \a\lex)\lex$ as well-founded ordering (H2).
  
  If the reduction does not take place at the root, then $w =
  f(\vec{u'})$ with $\vec{u} \a\lex \vec{u'}$. Since computability
  predicates are stable by reduction, the terms in $\vec{u'}$ are
  computable. Now, it is not difficult to see that $\succ$ is
  compatible with $\a$, that is, $u \succeq u'$ whenever $u\a u'$.
  Hence, by induction hypothesis (H2), $w$ is computable.
  
  If the reduction takes place at the root, then there are a rule
  $f(\vec{l}) \a r$ and a substitution $\t$ such that $dom(\t) =
  FV(\vec{l})$, $\vec{u} = \vec{l}\t$ and $w = r\t$. By definition of
  the General Schema, every variable $x$ free in $r$ is accessible in
  $\vec{l}$. Hence, by Lemma~\ref{lem-comp-acc-red} (take the identity
  for $\t'$), for all $x \in FV(r)$, $x\t$ is computable since, by
  hypothesis, the terms in $\vec{l}\t = \vec{u}$ are computable.
  Therefore the substitution $\t|_{FV(r)}$ is computable.
  
  We now show by induction on $r \in \C_f(\vec{l})$ that, for any
  computable substitution $\t'$ such that $dom(\t') \cap FV(r) =
  \emptyset$, $r\t\t'$ is computable (H3).

\begin{enumerate}
\item $r$ is a variable $x$. If $x \in dom(\t\t')$ then $r\t\t' =
  x\t\t'$ is computable since $\t\t'$ is computable. If $x \notin
  dom(\t\t')$ then $r\t\t' = x$ is computable since any variable is
  computable.
  
\item $r \in Acc(\vec{l})$. By Lemma~\ref{lem-comp-acc-red}.
  
\item $r = (v ~w)$ with $v$ and $w$ in $\C_f(\vec{l})$. By induction
  hypothesis (H3), $v\t\t'$ and $w\t\t'$ are computable. Therefore, by
  definition of computability predicates, $r\t\t'$ is computable.
  
\item $r = \l x. v$ with $v \in \C_f(\vec{l})$. Let $s \a t$ be the
  type of $r$ and $w$ be a computable term of type $s$. By induction
  hypothesis (H3), $v\t\t' \{x\mapsto w\}$ is computable. Hence, by
  Lemma~\ref{lem-comp-prop}.4, $r\t\t'$ is computable.
  
\item $r = g(\vec{v})$ with $g <_\cF f$ and each $v_i \in
  \C_f(\vec{l})$. By induction hypothesis (H3), each $v_i\t\t'$ is
  computable. Hence, by induction hypothesis (H1), $r\t\t'$ is
  computable.
  
\item $r = g(\vec{v})$ with $g =_\cF f$, each $v_i \in \C_f(\vec{l})$
  and $\vec{l} >\stat \vec{v}$. By induction hypothesis (H3), each
  $v_i\t\t'$ is computable. We show that $\vec{l}\t\t' \succ\stat
  \vec{v}\t\t'$.

\begin{itemize}
\item Assume that $l_i > v_j$ and $l_i$ is of type a strictly positive
  inductive type $\ts$. By definition of $>$, there is $p\in Pos(l_i)$
  such that $p\neq\vep$, $v_j = (l_i|_p ~\vec{v})$ and, for all $q<p$,
  $l_i|_q$ is constructor-headed. By assumption, each $v_i$ belongs to
  the computable closure. So, by induction hypothesis (H3), $v_i\t\t'$
  is computable. Now, $l_i|_p$ has a type of the form $\vec{s}\a\ts$.
  Let $\ts_q$ be the type of $l_i|_q$. Since $>_\cI$ is well-founded,
  all the $\ts_q$'s are equivalent to $\ts$. Thus, if $p=i_1\ldots
  i_{k+1}$ then $l_i\t\t' \succ l_i|_{i_1}\t\t' \succ \ldots
  l_i|_{i_1\ldots i_k}\t\t' \succ (l_i|_p\t\t' ~\vec{v}\t\t')$.
  
\item $v_j$ is a strict subterm of $l_i$ such that $FV(v_j) \sle
  FV(l_i)$. Hence, $v_j\t\t'$ is a strict subterm of $l_j\t\t'$ and
  $v_j\t\t' \succ l_j\t\t'$.
\end{itemize}
  
  Therefore, by induction hypothesis (H2), $r\t\t'$ is computable.
\end{enumerate}
\end{proof}


We are now able to prove the main lemma for strong normalization, i.e.
every term is computable. The strong normalization itself will follow
as a simple corollary.

\begin{lem}[Main lemma]
\label{lem-main-for-sn}
Assume that all the rules of $R$ follow the General Schema. Then, for
every term $u$ and computable substitution $\t$, $u\t$ is computable.
\end{lem}

\begin{proof}
We proceed by induction on the structure of $u$.

\begin{enumerate}
\item $u$ is a variable $x$. If $x \in dom(\t)$ then $u\t = x\t$ is
  computable since $\t$ is computable. If $x \notin dom(\t)$ then $u\t
  = x$ is computable since any variable is computable.
  
\item $u = f(\vec{u})$. By induction hypothesis, each $v_i\t$ is
  computable. Therefore, by Lemma~\ref{lem-comp-hofs}, $u\t$ is
  computable.
  
\item $u = \l x. v$. Let $s \a t$ be the type of $u$, $w$ be a
  computable term of type $s$ and $\t' = \t \uplus \{ x \to w \}$. By
  induction hypothesis, $v\t'$ is computable. Therefore, by
  Lemma~\ref{lem-comp-prop}.4, $(u\t ~w)$ is computable and $u\t$
  also.
  
\item $u = (v ~w)$. By induction hypothesis, $v\t$ and $w\t$ are
  computable. Therefore, by definition of computability, $u\t$ is
  computable.
\end{enumerate}
\end{proof}


\begin{thm}[Strong normalization]
\label{th-sn}
Under the assumptions 1 and 2, the combination of

\begin{enumerate}
\item the simply-typed $\l$-calculus with $\b\eta$-reductions and
\item higher-order rewrite rules following the General Schema
\end{enumerate}

is strongly normalizing.
\end{thm}

\begin{proof}
  Since a computability predicate of type $s$ contains all variables
  of type $s$, the identity substitution is computable. Hence, by
  Lemma~\ref{lem-main-for-sn}, every term is computable. And since
  computable terms are strongly normalizable, every term is strongly
  normalizable.
\end{proof}



\section{Examples and Extensions}
\label{sec-applications}

In this section, we present several applications and current
limitations of the General Schema termination proof method.


\subsection{Recursors for strictly positive types}

We already saw that the addition on Brouwer's ordinals follows the
General Schema. This is also true of the recursor on Brouwer's
ordinals \cite{stenlund72book}, as the user can easily check it:

\begin{rewc}
  $ordrec_t(X, Y, Z, 0)$ & $X$\\
  $ordrec_t(X, Y, Z, s(n))$ & $(Y ~n ~ordrec_t(X, Y, Z, n))$\\
  $ordrec_t(X, Y, Z, lim(F))$ & $(Z ~F ~\l n. ordrec_t(X, Y, Z, (F~ n)))$\\
\end{rewc}

where $ordrec_t$ is of type $t \a (\mt{ord} \a t \a t) \a ((\mt{nat}
\a \mt{ord}) \a (\mt{nat} \a t) \a t) \a \mt{ord} \a t$.

This is true as well of the recursors on mutually inductive types,
such as the type for trees:

\begin{center}
\begin{tabular}{rcl}
  $treerec_t(X, Y, Z, node(l))$ & $~\a~$ &
              $(X ~l ~listtreerec_t(X, Y, Z, l))$\\
  $listtreerec_t(X, Y, Z, nil)$ & $~\a~$ & $Y$\\
  $listtreerec_t(X, Y, Z, cons(x, l))$ & $~\a~$\\
  & & \hspace*{-3cm}
  $(Z ~x ~l ~treerec_t(X, Y, Z, x) ~listtreerec_t(X, Y, Z, l))$\\
\end{tabular}
\end{center}

The same property is actually true of arbitrary strictly positive
inductive types. The general case is no more difficult apart for the
more complex notations.

The uniqueness rules for recursors of basic inductive types were
studied in \cite{okada99tac} and extended to the strictly positive
case in \cite{hasebe00master}. In both cases, the termination proof
did not use the General Schema since the uniqueness rules do not seem
to fit the General Schema. It is open whether one could modify the
schema to cover this kind of rules.


\subsection{Curried function symbols}

We have assumed that all function symbols come along with all their
arguments. This is due to the fact that $\eta$ together with rewrite
rules over curried symbols lead to non-confluence. Take for example
$id : \mt{nat} \a \mt{nat}$ defined by $(id ~x) \a x$. Then, $\l x. x
\al \l x. (id ~x) \a_\eta id$.

Using curried symbols, however, is possible to the price of
duplicating the vocabulary as follows: for each function symbol $f$ of
arity $n>0$, we add a new function symbol $f^c$ of the same type as
$f$ but of arity $0$, defined by the rule

\begin{rewc}
$f^c$ & $\l x_1 \ldots \l x_n. f(x_1, \ldots, x_n)$
\end{rewc}

which satisfies the General Schema. Here is an example of definition
of the sum of a list of natural numbers using the $\mi{foldl}$ function:

\begin{rewc}
$\mi{foldl}(F, x, nil)$ & $x$\\
$\mi{foldl}(F, x, cons(y, l))$ & $\mi{foldl}(F, (F ~x ~y), l)$\\
$+^c$ & $\l x y . x + y$\\
$sum(l)$ & $\mi{foldl}(+^c, 0, l)$\\
\end{rewc}


\subsection{First-order rewriting}

In \cite{jouannaud97tcs}, the last two authors proved that it was
possible to combine higher-order rewrite rules following the General
Schema with a first-order rewrite system whose rules decrease in some
rewrite ordering and are non-duplicating (ie. no free variable occurs
more often in the right-hand side than in the left-hand side), a
condition needed to avoid Toyama's counter-example to the modularity
of termination \cite{toyama86}. It is of course possible to do the
same here, using Lemma 24 of \cite{jouannaud97tcs}, an analog of
Lemma~\ref{lem-comp-hofs} for first-order functions symbols.

Below, we give an example which cannot be proved to terminate by our
method: let $-$ and $/$ be the subtraction and division over natural
numbers. Note that $-$ follows the General Schema while $/$ does not
and that the last rule is duplicating the variable $y$:

\begin{center}
\begin{tabular}{l@{\hspace{1.5cm}}r}

\begin{rew}
$0 - y$ & $0$\\
$s(x) - 0$ & $s(x)$\\
$s(x) - s(y)$ & $x - y$\\
\end{rew}

& \begin{rew}
$x ~/~ 0$ & $x$\\
$0 ~/~ s(y)$ & $0$\\
$s(x) ~/~ s(y)$ & $s((x - y) ~/~ s(y))$\\
\end{rew}\\

\end{tabular}
\end{center}

In \cite{gimenez98icalp}, Gim\'enez proposes a terminating schema
using a notion of subtyping which allows to prove the strong
normalization property of this example.

However, we do not think this is a real issue. Non-termination does
not necessarily imply logical inconsistence, i.e. {\em False} is
provable. In the case of Toyama's counterexample to the modularity
property of termination, the union of the two original confluent and
terminating rewrite systems is not terminating, but every term has a
computable normal form. We believe, hence conjecture, that this
property is enough here to ensure that {\em False} cannot be derived
in the combined calculus.


\subsection{Conditional rewriting}

A conditional rule is a triple written $(l \a r ~\mbox{if}~ C)$ where
$C$ is a {\em condition} of the form $u_1 = v_1 \et \ldots \et u_n =
v_n$ with $FV(C) \sle FV(l)$, meaning that $l \a r$ may be applied
only if the terms of each pair $(u_i, v_i)$ have a common reduct. The
conditional rule:

\begin{center}
$l \a r ~~\mr{if}~~ u_1 = v_1 \et \ldots \et u_n = v_n$
\end{center}

can be encoded with the two non-conditional rules:

\begin{rewc}
$l$ & $eq_n(u_1, v_1, \ldots, u_n, v_n, r)$\\
$eq_n(x_1, x_1, \ldots, x_n, x_n, z)$ & $z$\\
\end{rewc}

The second rule satisfies the General Schema quite trivially. We
therefore say that a conditional rule follows the General Schema if $l
\a r$ follows the General Schema and $u_1, v_1, \ldots, u_n, v_n$ are
all in the computable closure of $l$. Hence, after
Theorem~\ref{th-sn}, if all the conditional rules satisfy the General
Schema, then $\a \cup \a_{\b\eta}$ is strongly normalizing.

A well-known example is given by an insertion function on lists.

\begin{rewc}
$insert(x, nil)$ & $cons(x, nil)$\\
$insert(x, cons(y, l))$ & $cons(x, cons(y, l))$
~if~ $\mi{inf}(x,y) = true$\\
$insert(x, cons(y, l))$ & $cons(y, insert(x, l))$
~if~ $\mi{inf}(x,y) = false$\\
$\mi{inf}(0,x)$ & $true$\\
$\mi{inf}(s(x),0)$ & $false$\\
$\mi{inf}(s(x),s(y))$ & $\mi{inf}(x,y)$\\
\end{rewc}


\subsection{Congruent types}

We are going to see that our method can easily cope with basic
inductive types whose constructors satisfy some (first-order)
equations, provided that these equations form a weakly-normalizing
term rewriting system, that is, such that every term has a unique
normal form. In this case, the initial algebra of the inductive type
is equivalent to its normal form algebra and the latter can be
represented by the accepting states of a finite tree automaton of some
form \cite{bouhoula99tcs,comon97book}. The important property of this
automaton is that the set of terms recognized at every accepting state
is recursive and the predicate of this state is actually easy to
define. We show the construction for the simple example of integers.
The general case of an arbitrary basic inductive type is no different.

\begin{figure}[ht]
\begin{center}
\scalebox{.7}{\includegraphics{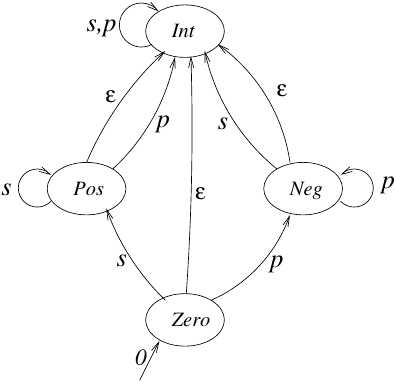}}
\caption{Automaton\label{fig-automaton}}
\end{center}
\end{figure}

The inductive type $\mt{int}$ is specified with the constructors $0$,
$s$ and $p$ for zero, successor and predecessor respectively, and the
two equations: $s(p(x)) = x$ and $p(s(x)) = x$, which are easily
turned into a first-order convergent term rewriting system $\{s(p(x))
\a x, ~p(s(x)) \a x\}$ whose normal forms are recognized by the
automaton given at Figure~\ref{fig-automaton}. This automaton can be
easily constructed by solving disequations over terms (see
\cite{comon97book,monate97dea}).

Then, the recursor on integers may be defined by the following set of
constraint rules:

\begin{rewc}
$intrec_t(X, Y, Z, 0)$ & $X$\\
$intrec_t(X, Y, Z, s(x))$ & $(Y ~x ~intrec(X, Y, Z, x))$ ~if~ $s(x) \in Pos$\\
$intrec_t(X, Y, Z, p(x))$ & $(Z ~x ~intrec(X, Y, Z, x))$ ~if~ $p(x) \in Neg$\\
\end{rewc}

As usual, it is then possible to define other functions such as the
addition by the use of the recursor:

\begin{rewc}
$x + y$ & $intrec_\mt{int}(x, \l xy. s(y), \l xy. p(y), y)$
\end{rewc}

which is equivalent to the following pattern-matching definition:

\begin{rewc}
$x + 0$ & $x$\\
$x + s(y)$ & $s(x + y)$ ~if~ $s(y) \in Pos$\\
$x + p(y)$ & $p(x + y)$ ~if~ $p(y) \in Neg$\\
\end{rewc}

but to which we may add, for example, the rule for associativity:

\begin{rewc}
$(x + y) + z$ & $x + (y + z)$\\
\end{rewc}

or, by a completely different definition which does not make use of
the automaton but makes use of the signature present in the user's
specification only:

\begin{center}
\begin{tabular}{l@{\hspace{1.5cm}}r}

\begin{rew}
$x + 0$ & $x$\\
$x + s(y)$ & $s(x + y)$\\
$x + p(y)$ & $p(x + y)$\\
\end{rew}

& \begin{rew}
$s(p(x))$ & $x$\\
$p(s(x))$ & $x$\\
\end{rew}\\

\end{tabular}
\end{center}

It is of course a matter of debate whether the normal form
computations should be made available to the users, like the
recursors, or should not. We have no definite argument in favor of
either alternative.

We have assumed that the specification of constructors was a weakly
normalizing (in practice, a confluent and terminating set) of rewrite
rules. The method applies as well when some constructor is commutative
or, associative and commutative (with some additional technical
restriction). See \cite{bouhoula99tcs} for more explanations and
additional references. Whether it can be generalized to non-basic
inductive types is however open.


\subsection{Matching modulo $\b\eta$}

In this section, we address the case of higher-order rewrite rules
{\em \`a la} Nipkow \cite{nipkow91lics}, based on higher-order
pattern-matching with patterns {\em \`a la} Miller \cite{miller89elp}.
We give here several examples taken from \cite{nipkow91lics},
\cite{vandepol93hoa} or \cite{prehofer95thesis}, and recall why plain
pattern-matching does not really make sense for them. On the other
hand, we will see that all these examples follow the General Schema:
we explain the first example in detail and the user is invited to
check the others against our definitions.

We start with the example of differentiation of functions over the
inductive type $\mt{R}$:

\begin{center}
\begin{tabular}{cccc}

\begin{rew}
$x \times 1$ & $x$\\
$1 \times x$ & $x$\\
\end{rew}

& \begin{rew}
$x \times 0$ & $0$\\
$0 \times x$ & $0$\\
\end{rew}

& \begin{rew}
$x + 0$ & $x$\\
$0 + x$ & $x$\\
\end{rew}

& \begin{rew}
$0 ~/~ x$ & $0$\\
\end{rew}\\

\end{tabular}
\end{center}

\begin{rewc}
$D(\l x. y)$ & $\l x.0$\\
$D(\l x. x)$ & $\l x.1$\\
$D(\l x. sin(F ~x))$ & $\l x. cos(F ~x) \times (D(F) ~x)$\\
$D(\l x. cos(F ~x))$ & $\l x. - sin(F ~x) \times (D(F) ~x)$\\
$D(\l x. (F ~x)+(G ~x))$ & $\l x. (D(F) ~x) + (D(G) ~x)$\\
$D(\l x. (F ~x) \times (G ~x))$ & $\l x.(D(F) ~x) \times (G ~x) +
(F ~x) \times (D(G) ~x)$\\
$D(\l x. ln(F ~x))$ & $\l x. (D(F) ~x) ~/~ (F ~x)$\\
\end{rewc}

Note first that we cannot have composition explicitly as a constructor
of the inductive type $\mt{R}$, since the positivity condition would
be violated. We could define it with the rule $F \circ\, G \a \l x. (F
~(G ~x))$, but then, in $D(F \circ\, G)$, $F$ and $G$ are not
accessible since they are not of basic type and, in $D(\l x. (F ~(G
~x)))$, $F$ is not accessible since it is not applied to distinct
bound variables, a condition also required for patterns in Nipkow's
framework. This explains why composition is encoded in each rule by
using the application operator of the $\l$-calculus.

The rules defining $\times$, $+$ and $/$ are usual first-order rules.
We could restrict the use of the last one to the case where $x$ is
different from $0$. Of course, this is not possible with a faithful
axiomatization of reals, since equality to $0$ is not decidable for
the reals. As for the other rules, $D(\l x. y) \a \l x.0$ states that
the differential of a constant function (equal to $y$) is the null
function. The definition of substitution ensures here that $x$ cannot
occur freely in an instance of $y$, hence $y$ is a constant with
respect to $x$ (although it may depend on other variables free in the
rewritten term). The rule $D(\l x. x) \a \l x.1$ states that the
differential of the identity is the constant function equal to $1$.
The next rule, $D(\l x.sin(F ~x)) \a \l x.  cos(F ~x) \times (D(F)
~x)$, defines the differential of a function obtained by composing
$sin$ with some other function $F$. The other rules speak for
themselves.

Assume now that we use first-order pattern-matching for these rules.
Then, we would not be able to differentiate the function $\l x .
sin(x)$ by computing $D(\l x . sin(x))$, because no rule would match.
Of course, we could give new rules for this case, but this would be an
endless game. The use of higher-order matching, on the other hand,
chooses the appropriate value for the higher-order free variables so
as to cover all cases.

The local confluence of these rules can be checked on higher-order
critical pairs, as shown by Nipkow \cite{nipkow91lics,mayr98tcs}. The
computation of these critical pairs can be done in linear time
\cite{qian93tapsoft}, thanks to the hypothesis that the left-hand
sides are patterns.

We now show that this example follows the General Schema, by showing
first that the free variables of the right-hand sides are accessible
in their respective left-hand side. For the rule $D(\l x. y) \a \l x.
0$, $y$ is accessible in $\l x. y$ by cases (1) and (2). For the rule
$D(\l x. sin(F ~x)) \a \l x. cos(F ~x) \times (D(F) ~x)$, $F$ is
accessible in $\l x. sin(F ~x)$ by (1), (2), (3) and (4). Now, it is
not difficult to check that the right-hand sides belong to the
computable closure of their respective left-hand side.

Prehofer and van de Pol prove the termination of this system (with
higher-order pattern-matching) by defining a higher-order
interpretation proved to be strictly monotonic on the positive natural
numbers \cite{prehofer95thesis}, a method developed by van de Pol
\cite{vandepol93hoa} that generalizes to the higher-order case the
interpretation method of first-order term rewriting systems. One can
easily imagine that it is not easy at all to find higher-order
interpretations. Here, $D$ needs to be interpreted by a functional
which takes as arguments a function $f$ on positive natural numbers
and a positive natural number $n$, for example the function $(f,n) \to
1 + n \times f(n)^2$. Furthermore, the interpretation method is not
modular, the adequate interpretation of each single function symbol
depending on the whole set of rules. This makes it difficult to use by
non-experts.

The next example is taken from process algebra \cite{sellink93sosl}:

\begin{rewc}
$p + p$ & $p$\\
$(p + q) \,; r$ & $(p \,; q) + (q \,; r)$\\
$(p \,; q) \,; r$ & $p \,; (q \,; r)$\\
$p + \delta$ & $p$\\
$\delta \,; p$ & $\delta$\\
$\S(\l d. p)$ & $p$\\
$\S(X) + (X ~d)$ & $\S(X)$\\
$\S(\l d. (X ~d) + (Y ~d))$ & $\S(X) + \S(Y)$\\
$\S(X) \,; p$ & $\S(\l d. (X ~d)\,; p)$\\
\end{rewc}

Note that the left-hand side of rule $\S(X) + (X ~d)$ is not a pattern
{\em \`a la} Miller. As a consequence, Nipkow's results for proving
local confluence do not apply. Termination of these rules is also
proved in \cite{vandepol93hoa}. To see that this example follows the
General Schema, it suffices to take the precedence defined by $; >
\delta, \S$ and $\S > +$. The rule $\S(X) + (X ~d) \a \S(X)$, which is
a simple projection, is dealt with by case (2).

The last example, the computation of the negative normal form of a
formula, is taken from logic (we give only a sample of the rules):

\begin{rewc}
$\neg(\neg(p))$ & $p$\\
$\neg(p \et q)$ & $\neg(p) \ou \neg(q)$\\
$\neg(\all(P))$ & $\exists(\l x. \neg(P ~x))$\\
\end{rewc}

Of course, the fact that all the above examples follow the General
Schema does not imply that Nipkow's rewriting terminates. However, we
conjecture that it does and that it is due to the use of patterns in
the left-hand sides. To prove our conjecture, we essentially need to
show that higher-order pattern-matching preserves computability. This
has been recently proved by the first author in \cite{blanqui00rta},
where the framework described here is extended into a typed version of
Klop's higher-order rewriting framework \cite{klop93tcs}, and where
Nipkow's higher-order Critical Pair Lemma is shown to apply to this
extended framework also.


\subsection{Rewriting modulo additional theories}

It is of general practice to rewrite modulo properties of constructors
(implying that the underlying inductive type is a quotient) or defined
symbols. Usual properties, as in presentations of arithmetic, are
commutativity or, commutativity and associativity. In our encoding of
predicate calculus, there is a less common kind of commutativity of
bound variables, expressed by the equation:

\begin{center}
$\all(\l x . \all(\l y . (P ~x ~y))) =
\all(\l y . \all(\l x . (P ~x ~y)))$
\end{center}

We now give (a sample of) the rules for the computation of the prenex
normal form of a formula:

\begin{rewc}
$\all(P) \et q$ & $\all(\l x. (P ~x) \et q)$\\
$p \et \all(Q)$ & $\all(\l x. p \et (Q ~x))$\\
\end{rewc}

The above set of rules is confluent modulo the previous equation (but
would not be confluent directly). Note that matching modulo the
equation is not necessary here because of the form of the left-hand
sides of rules.

We end this list with ``miniscoping'', an operation inverse of the
prenex normal form:

\begin{rewc}
$\all(\l x. p)$ & $p$\\
$\all(\l x. (P ~x) \et (Q ~x))$ & $\all(P) \et \all(Q)$\\
$\all(\l x. (P ~x) \ou q)$ & $\all(P) \ou q$\\
$\all(\l x. p \ou (Q ~x))$ & $p \ou \all(Q)$\\
\end{rewc}

These examples follow our schema as well. Of course, this does not
prove strong normalization, since we did not prove that the schema is
compatible with such theories. The generalization is quite
straightforward for commutativity but needs more investigations for
more complex theories such as the above one or, associativity and
commutativity together.



\section{Conclusion}
\label{sec-conclusion}

This paper is a continuation of \cite{jouannaud97tcs}. Our most
important contributions are the following:

\begin{enumerate}
  
\item Our new General Schema is strong enough so as to capture
  strictly positive recursors, such as the recursor for Brouwer's
  ordinals, without compromising the essential properties of the
  calculus. The strong normalization proof for this extension is again
  based on the Tait and Girard's computability predicates technique
  and uses in an essential way the strict-positivity condition of the
  inductive types.

\item The new formulation of the schema makes it very easy to define
  new extensions, by simply adding new cases to the definition of
  ``accessibility'', or new computability preserving operations in the
  ``computable closure''.
  
\item The notion of ``computable closure'' is an important concept
  which has already be used in a different context
  \cite{jouannaud99lics}.
  
\item Several precise conjectures have been stated. The most important
  two, in our view, are the use of the General Schema to prove the
  strong normalization of higher-order rewriting {\em \`a la} Nipkow
  on the practical side, and the generalization of the schema to
  capture (non-strictly) positive inductive types on the theoretical
  one. The first conjecture has been recently solved by the first
  author in \cite{blanqui00rta}.
\end{enumerate}

Another kind of extension should now be considered, by considering a
richer type system, which we did in \cite{blanqui99rta}, keeping the
same definition for the rules and the General Schema. But a richer
type system allows us to have richer forms of rewrite rules: the
General Schema should therefore be adapted so as to allow for rules of
a dependent type and even rules over types. Experience shows that the
latter kind of extension raises important technical difficulties.
Strong elimination rules in the Calculus of Inductive Constructions
\cite{werner94thesis} or the rules defining a system of Natural
Deduction Modulo \cite{dowek98tr3400,dowek98tr3542} are of that kind.




\end{document}